\documentclass[12pt,preprint]{aastex}

\newcommand\ergscmA{\ifmmode {\rm ergs\,s}^{-1}\,{\rm cm}^{-2}\,{\rm \AA}^{-1}
	  \else ergs\,s$^{-1}$\,cm$^{-2}$\,\AA$^{-1}$\fi}
\newcommand\cc{\ifmmode {\rm cm}^{-3} \else cm$^{-3}$\fi}
\newcommand\cl{\ifmmode {\rm cm}^{-2} \else cm$^{-2}$\fi}
\newcommand\kms{\ifmmode {\rm km\,s}^{-1} \else km\,s$^{-1}$\fi}

\newcommand\gtsim{\raisebox{-.5 ex}{$\;\stackrel{>}{\sim}\;$}}
\newcommand {\HST}{{\it HST}}
\newcommand {\lya}{Ly$\alpha$}

\newcommand {\he}{He~{\sc ii}}
\newcommand {\h}{H~{\sc i}}
\newcommand {\lm}{$\lambda$}
\newcommand {\etal}{et al.\ }
\slugcomment{To appear in {\it The Astronomical Journal}} %
\shorttitle{Intergalactic helium}
\shortauthors{Zheng et al.}

\begin{document}

\title{Detection of Intergalactic He {\sc ii} Absorption at Redshift 
3.5\altaffilmark{1}}

\author{
Wei Zheng\altaffilmark{2}, 
Kuenley Chiu\altaffilmark{2}, 
Scott F. Anderson\altaffilmark{3},
Donald P. Schneider\altaffilmark{4},
Craig J. Hogan\altaffilmark{3},
\\
Donald G. York\altaffilmark{5},
Scott Burles\altaffilmark{6}
and
Jonathan V. Brinkmann\altaffilmark{7}
}

\altaffiltext{1}{
Based on observations made with the NASA/ESA Hubble Space Telescope, obtained
at the Space Telescope Science Institute, which is 
operated by the Association of Universities for Research in Astronomy, Inc., 
under NASA contract NAS 5-26555. These observations are associated with 
program GO 8582 and 9067.
}
\altaffiltext{2}{Department of Physics and Astronomy, Johns Hopkins 
University, Baltimore, MD 21218} 
\altaffiltext{3}{Department of Astronomy, University of Washington, Seattle, 
WA 98195}
\altaffiltext{4}{Department of Astronomy and Astrophysics, Pennsylvania State 
University, University Park, PA 16802} 
\altaffiltext{5}{Department of Astronomy, University of Chicago, Chicago, 
IL 60637}
\altaffiltext{6}{Department of Physics, Massachusetts Institute of Technology,
Cambridge, MA 02139}
\altaffiltext{7}{Apache Point Observatory, New Mexico State University, 
 Sunspot, NM 88349}
\begin{abstract}
The large number of quasars found in the Sloan Digital Sky Survey has 
allowed searches for elusive, clear lines of sight towards 
He {\sc ii} Ly$\alpha$ absorption,
a sensitive probe of the intergalactic medium.
The few known systems indicate that He {\sc ii} reionization occurs 
at $z >3$. We report the detection of a He {\sc ii} Ly$\alpha$ 
absorption edge in a quasar spectrum at $z=3.50$, the most distant such  
feature found to date.
The candidate quasar was selected from a $z\sim3$ sample 
in the SDSS spectroscopic quasar survey and confirmed as part of an {\it
HST}/STIS
SNAP survey. We discuss the general characteristics of the absorption 
feature, as well as the probability for discovery of additional such objects. 
\end{abstract}

\section{INTRODUCTION}
Much of the intergalactic hydrogen 
exists in a highly ionized state as evidenced by the lack of \h\
Gunn-Peterson troughs in the optical spectra of high-redshift quasars 
\citep{gp}.
The first presence of a Gunn-Peterson trough was detected in \he\
absorption, in a quasar at $z = 3.28$
\citep{jakobsen}. Follow-up observations at a higher S/N level and spectral 
resolution \citep{hogan,heap} 
found an average \he\ optical depth $\tau = 4.0 \pm 0.5$ at $z \sim 3$ and
at least $\tau=5$ at $z\sim 3.2$. 
\he\ absorption displays significant structure, consisting of 
unabsorbed void regions and deep absorption troughs \citep{reimers}.
When contrasted with the low \he\ opacity 
measured at $z \sim 2.5$ \citep{dkz,gak,zheng} 
these findings indicate a sudden increase in optical depth. 
While the large opacity at $z \sim 3.2$ is conceivably related to a
fluctuation in the number of local ionizing sources, it may instead
(e.g., if confirmed in multiple additional independent sightlines) more
generally signal an epoch in which intergalactic helium is not fully ionized. 
Recently, a similar pattern for intergalactic hydrogen has been observed in 
the $z > 6$ quasars. 
\citet{becker} and \citet{fan} report a \h\ optical depth of at least 
$\tau=15$ at $z\sim 6.2$, signaling the late stages of reionization of the
intergalactic medium (IGM). The recent {\it WMAP} results \citep{map} suggest 
that the IGM reionization may have started at $z \sim 17$, and there may have 
been a second stage of hydrogen reionization that took place at $z\sim 6$ 
\citep{cen}. 

The reionization of intergalactic hydrogen and helium are cosmic milestones. 
As the early universe cooled and evolved, neutral hydrogen atoms in the IGM 
gradually became bathed in the ionizing ultraviolet background radiation 
produced by the first stars and quasars. With increasing UV background 
radiation, the IGM eventually became fully ionized.
Atomic hydrogen and helium are affected by the same ionizing UV radiation 
in the distant universe -- but because the \he\ \lya\ transition is
related to the number of ionizing photons at 4 Rydbergs (generally smaller than
the number at 1 Rydberg), the the reionization of intergalactic helium is 
expected to occur later than hydrogen, approximately at $z\sim 3-5$ 
\citep{haardt,madau2,loeb}.  Indirect evidence \citep{theuns} suggests that 
the final stage of \he\ reionization may take place at $z \sim 3.2-3.4$,
but only a very small number of independent sightlines have thus far  
been identified that are suitable to probe \he\ even to $z\sim 3$. Moreover,
each such 
sightline is potentially susceptible to systematic bias; these sightlines are, 
by selection, the most unobscured of all known quasar sightlines to high 
redshift. It is therefore important to find additional sightlines (to average 
over possible systematics in global conclusions), and also to extend \he\
studies to higher redshift. Higher redshift sightlines are, of course,
potentially the most useful as each such object may provide a measure of \he\ 
absorption spanning the full redshift range likely to be most interest. 
Our knowledge of the intergalactic helium has been derived from only four lines
of sight, and only two are between $3.0 < z < 3.3$. More such examples are 
needed to establish the statistics and evolution of intergalactic helium at 
redshift $z>3$. The discovery of any bright, UV-unobstructed quasar in this 
redshift range is therefore interesting.

We report here the detection of a \he\ absorption feature in the spectrum 
of quasar SDSS J234625.67-001600.4 (hereafter abbreviated as SDSSJ2346-0016),
the highest redshift \he\ feature found to date.  
This new quasar sightline potentially
will allow a measure of \he\ absorption from $z=2.8$ to $z=3.5$
in a single object, efficiently sampling the full redshift range 
commonly suggested for the epoch of helium reionization. We describe
the method of selection, preliminary observations from {\it HST}/STIS, 
and the probability for further discovery of similar objects. 

\section{CANDIDATE SELECTION AND OBSERVATIONS}

At redshift $z \gtsim 2.8$ the \he\ \lya\ feature becomes observable with 
\HST. However,
the observed UV flux of most high-redshift quasars is strongly suppressed by
absorption from intergalactic \h\ along the line of sight. For objects at 
$z \sim 3$, attenuation by individual \lya\ forest lines is moderate at 
restframe wavelengths shortward of $\sim 600$~\AA, but Lyman-limit 
systems (LLSs) can remove significant -- and frequently most -- of the UV flux 
in high-redshift quasars \citep{moller}.
Not until recently have quasar catalogs provided sufficient numbers of bright, 
high-redshift candidates to begin to outweigh these difficulties.
To date extensive searches have revealed only four such quasar sightlines with
\he\ absorption features.  Such a sample is too small to confidently understand 
the possible systematic uncertainties in the resulting cosmological inferences.

The Sloan Digital Sky Survey \citep[SDSS;][]{sdss} plans to image a
quarter of the sky through five broadband filters \citep{fukug} and to
obtain spectra of approximately one million galaxies and 100,000 quasars.
The survey employs a dedicated 2.5m telescope at Apache Point Observatory
to obtain images \citep{gunn}, and automated pipelines 
\citep{pier,smith,stoughton} process the data. 
Objects were selected based on their broadband SDSS colors \citep{Hogg}. 
\citet{richards} describe the SDSS quasar selection algorithm.  
The spectrophotometric calibration includes scaling by wider aperture "smear"
 exposures to the photometry from the imaging data.
Details regarding the spectroscopic system and analysis pipelines may be found 
in \citet{stoughton} and \citet{blanton}.
The Early Data Release (EDR) Quasar Catalog contains 3814 quasars,
3000 discovered by the SDSS \citep{edr}. 
The redshift-magnitude distribution of the 302 EDR quasars at $z>2.5$ are 
shown in Figure 1.

The SDSS provides an excellent sample in which to search for quasar candidates 
with \he\ absorption features.  
We selected a sample of optically bright ($i <18.5$) quasars at redshift 
$z>2.9$
for an \HST\ snapshot survey to search for \he\ absorption edges in the 
UV. The snapshot program used the Space Telescope Imaging Spectrograph (STIS); only one out of the 24 candidates observed by the
programs exhibited flux at the expected location of \he\ absorption. 

The optical spectrum of SDSSJ2346-0016 ($i=17.8$) was processed using the 
standard SDSS 
spectroscopic data reduction pipeline, and is shown in Figure 2. 
The redshift is measured using the peak of several emission lines:
\ion{C}{4} \lm 1549,  \ion{O}{1} \lm 1304, \ion{C}{2} \lm 1334,
\ion{Si}{4}+\ion{O}{4} \lm 1400, and \lya, and the averaged value is
$z = 3.49 \pm 0.03$. The \lya\ forest lines set in at $5476.2$~\AA,
corresponding to $z=3.5047$.

The ultraviolet \HST/STIS spectroscopic snapshot of SDSSJ2346-0016 was 
performed on 
2002 May 18. The data were obtained with the G140L grating and a $52\arcsec 
\times 0.5\arcsec$ slit. covering the wavelength range of 1150 to 1700~\AA.
The spectral resolution was approximately 1~\AA, and the exposure time was 
600 seconds.
The \HST\ spectrum extraction was done manually, and 
processed using the standard STIS pipeline for wavelength and flux 
calibration.  Figure 3 shows the spectrum binned by 9 pixels ($\sim 2.7$~\AA). 
It was fit with a power-law continuum, a \lya\ emission line, and an 
absorption edge. The fitted absorption edge is at $1369 \pm 2 
$~\AA, with a longward continuum level at $f_\lambda \sim 10^{-16} \ 
$ \ergscmA.
No flux is detected at the wavelengths below 1360~\AA. Between 1330 and 
1360~\AA, the average flux is $(0.05 \pm 0.8) \times 10^{-17}$ \ergscmA, 
corresponding to an optical depth of $6 ^{+\infty}_{-3}$.

The match between the redshifts of \h\ \lya\ and the 
\he\ \lya\ features strongly suggests, at a 98\% level, that the
observed UV absorption edge is not merely coincidental. The 2\% probability
is based on the expected number of LLSs per redshift interval \citep{lombardi}.
If this observed break is a LLS, the restframe cutoff wavelength is at 920~\AA\
or longer, due to the combined effect of high-order Lyman-series lines.
Such a system would be at $z \sim 0.49$, and a pair of \ion{Mg}{2} lines may 
be expected around 4163.7/4174.4~\AA. Between 4160 and 4180~\AA\ there are 
three 
absorption lines in the SDSS spectrum: 4180.6~\AA (EW$ \sim 2.4$~\AA), 
4167.3~\AA\ (EW$\sim 1.5$~\AA) and 4161.9~\AA\ (EW$\sim 4.4$~\AA). If any of 
them is \ion{Mg}{2} 2803~\AA, a counterpart feature is expected approximately 
at the wavelength 10.5~\AA\ shortward. No such lines are identified within
2~\AA\ at the expected wavelength and within 30\% of the expected intensity 
(50\%).

\section{IMPLICATIONS}

The study of \he\ absorption in this new unobscured quasar sightline
(and others extending to similarly high redshifts)
will allow significant progress in IGM studies. For example, if
helium in the IGM is not fully ionized at $z=3.50$, the high \he\ 
opacity would result in a damping profile, which may be detected even at low 
spectral resolution \citep{me}. 
This profile is characterized by a redward wavelength shift of a trough
from the  expected position of \he\ \lya\ emission, on the order of 2-10~\AA. 
The limited 
signal-to-noise level of our detection spectrum is insufficient to accurately 
measure such a feature;
future observations with STIS or COS (Cosmic Origins Spectrograph)
with significantly longer exposures
may establish conclusive evidence. The diagnosis of such a feature may be 
complicated by the proximity effect \citep{madau}, and residual infall of 
matter into the deep potential well of the quasar host 
galaxy may also give rise to such absorption feature \citep{barkana}.
However, previous observations of \he\ absorption in other $z>3$ quasars 
suggest that these effect may not be always present, as high-opacity clouds
near the quasar may block the ionizing radiation from the quasar itself 
\citep{heap}.

Should future observations of this object reveal the proximity effect,
the data are useful to estimate the intensity of the metagalactic 
UV background radiation \citep{zheng,me} and the baryon density of the IGM 
\citep{hogan,anderson}. It is known that \he\ absorption at $z>3$ is 
represented by high opacity of $\tau > 3$.
The wavelength baseline between 1050 and 1370~\AA\ in a \HST\ spectrum will be 
the longest to reveal the ionization history of intergalactic helium. 

This detection strengthens our confidence in future discoveries of 
similar features at higher redshifts.
Model simulations of the IGM \citep{moller} suggest a success rate of 
approximately 7\% in finding a $z >3$ quasar whose flux at 
300~\AA\ (restframe) may be detected. This appears to be consistent with our 
detection rate of 4\% (one in 24 objects) and that of a similar 
program (\HST\ GO 8287, one detection in 26 objects). While the lines of sight 
towards objects at even higher redshifts will encounter more intervening 
LLSs,
these additional systems will not affect the \he\
detection in a significant manner -- they set in at long 
wavelengths ($\gtsim 4100$~\AA), and their optical depths recover quickly 
with $\tau \propto \lambda^3$.  Therefore, the effect on \he\ detection at 
$z>3.5$ is determined mainly by the LLSs at $z<3.5$. 
The large number of SDSS quasars should provide 
several hundreds of
quasars at $z>3.5$ that are sufficiently bright for future study. The 
cases confirmed to be unobscured will be studied with \HST's STIS and/or Cosmic
Origins Spectrograph (COS) with significantly higher sensitivity. 

The large span in IGM redshift accessible to \he\ studies
in each such suitable high redshift quasar will provide critical
information on the reionization history of the IGM.
The anticipated large sample size will allow assessment of,
and averaging over, potential systematic biases that may
be present in the current critically small sample, as well
as the cosmological inferences obtained therefrom.

\acknowledgments

Funding for the creation and distribution of the SDSS archive has been provided 
by the Alfred P. Sloan Foundation, the Participating Institutions, the National
Aeronautics and Space Administration, the National Science Foundation, the U.S. 
Department of Energy, the Japanese Monbukagakusho, and the Max Planck Society. 
The SDSS Web site is http://www.sdss.org/.

The SDSS is managed by the Astrophysical Research Consortium (ARC) for the 
Participating Institutions. The Participating Institutions are The University of 
Chicago, Fermilab, the Institute for Advanced Study, the Japan Participation 
Group, The Johns Hopkins University, Los Alamos National Laboratory, the 
Max-Planck-Institute for Astronomy (MPIA), the Max-Planck-Institute for 
Astrophysics (MPA), New Mexico State University, University of Pittsburgh, 
Princeton University, the United States Naval Observatory, and the University of 
Washington.

Support for this research was provided by NASA through grants 8582 and 9067
from the Space Telescope Science Institute, which is operated by the 
Association of Universities for Research in Astronomy, Inc., under NASA
contract NAS 5-26555.

\clearpage

\begin{figure}
\plotone{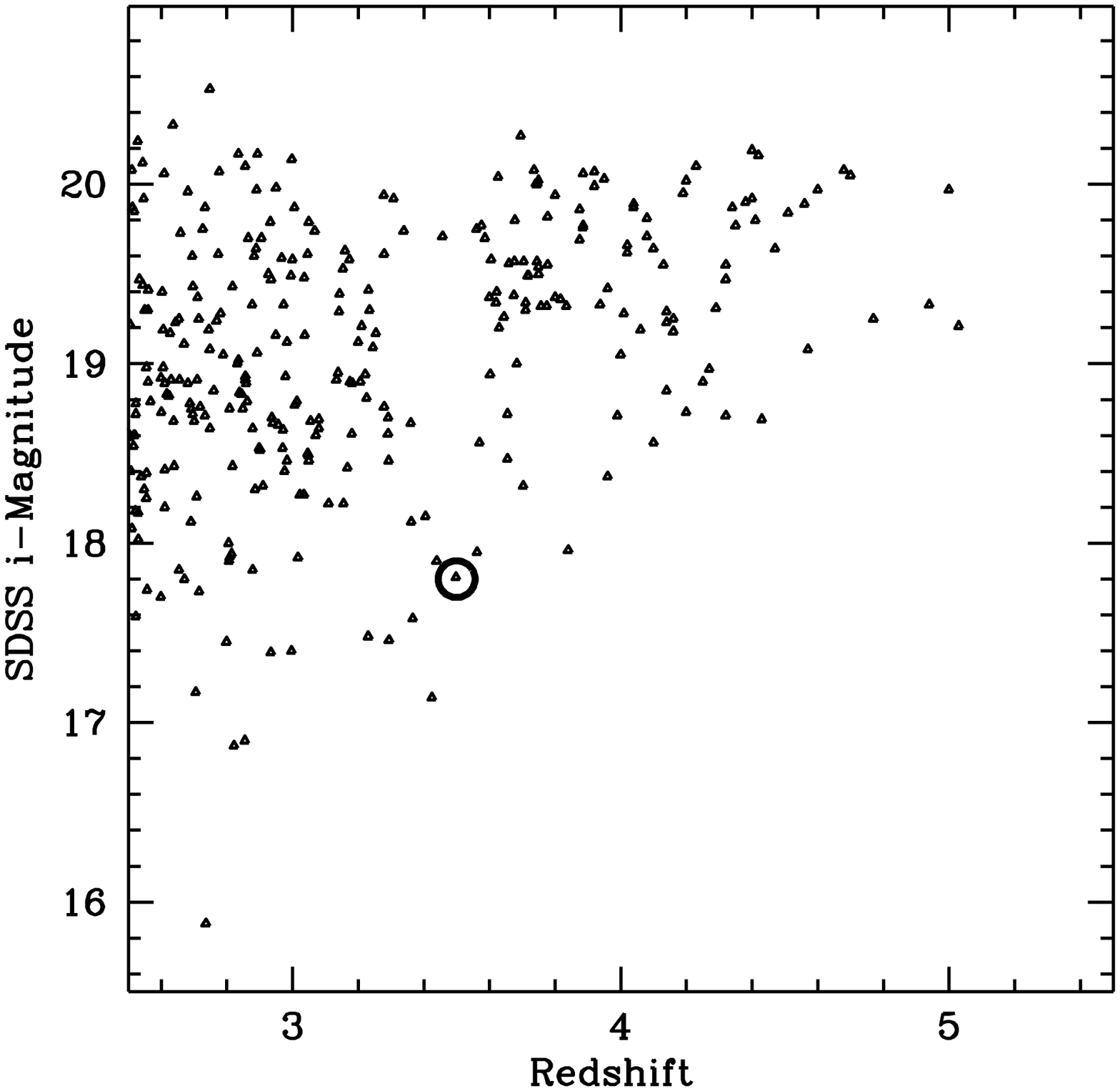}
\caption{ 
302 quasars at $z>2.5$ found in the SDSS EDR \citep{edr}. 
The circle marks the position for the newly discovered quasar SDSSJ2346-0016. 
\label{fig1}}
\end{figure}

\begin{figure}
\plotone{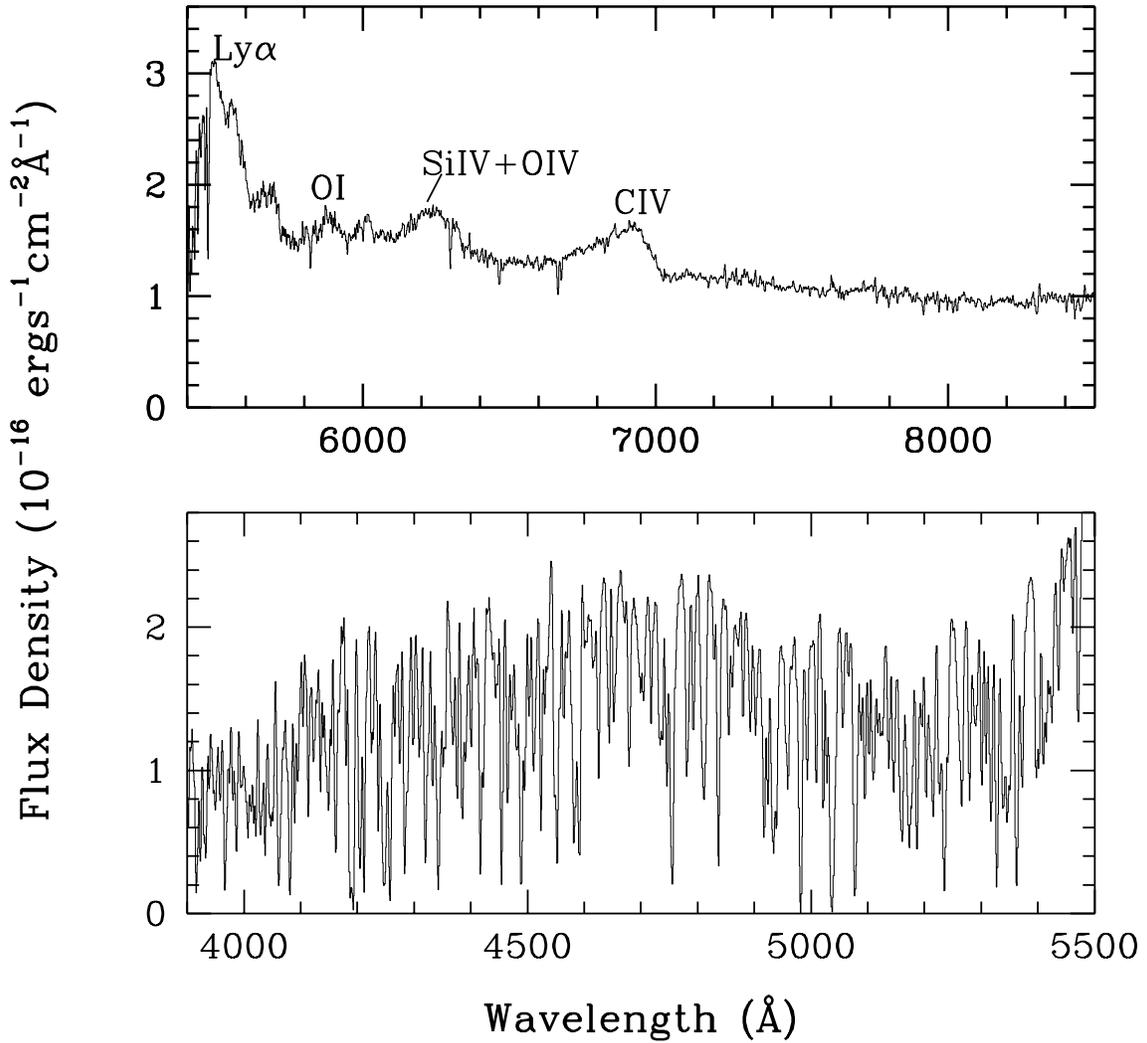}
\caption{
Optical spectrum of SDSSJ2346-0016.  The spectral resolution is approximately
1900. The top panel displays the spectral region longward of 5400~\AA, and
the bottom panel for the Ly$\alpha$ forest region. 
Major emission lines are indicated in the figure.
\label{fig2}}
\end{figure}

\begin{figure}
\plotone{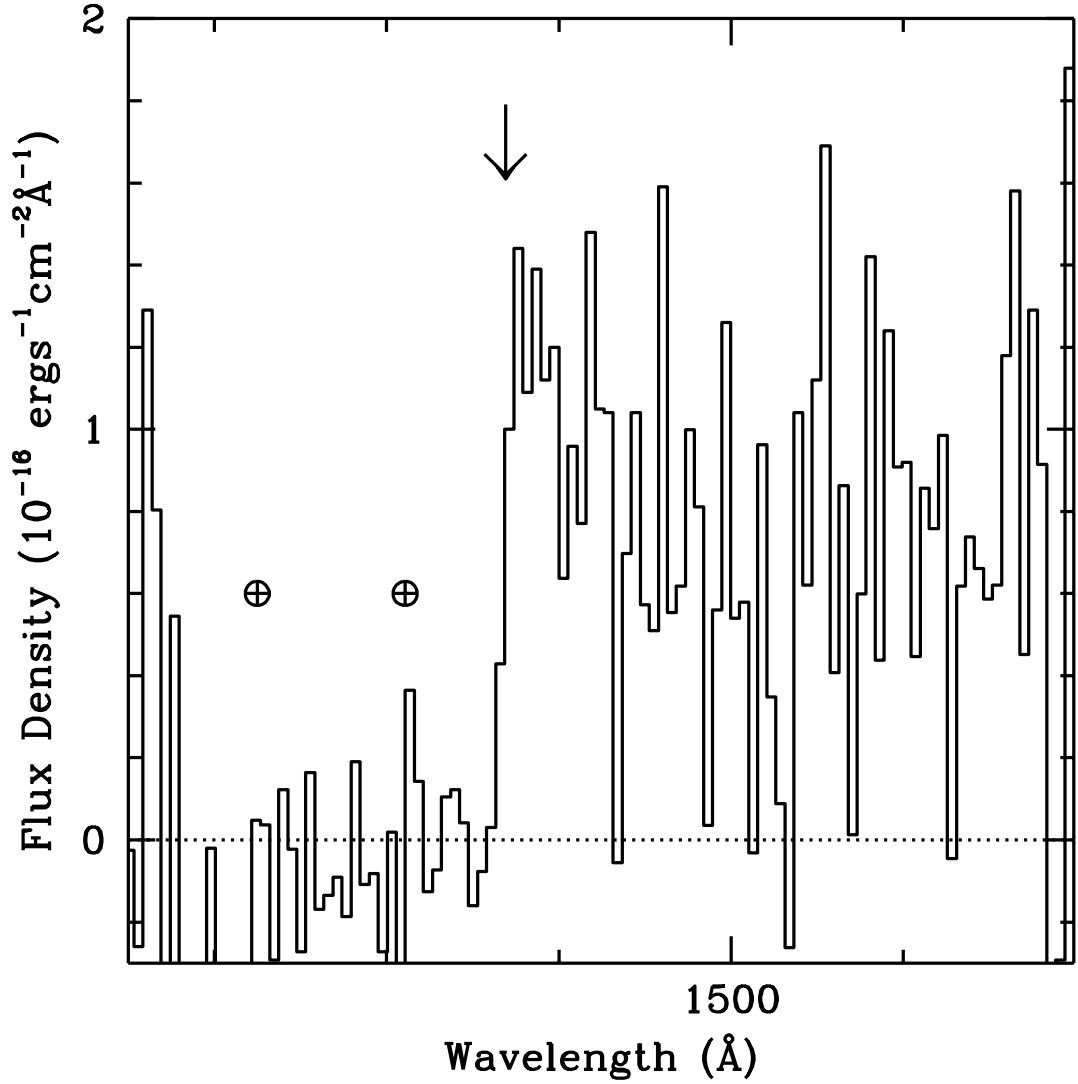}
\caption{
{\it HST}/STIS spectrum of quasar SDSSJ2346-0016. The G140L data are binned 
by nine pixels ($\sim 2.7$~\AA). The He {\sc ii} Ly$\alpha$ absorption edge 
is near the expected wavelength of 1370~\AA.  The Earth symbols mark the 
wavelengths of two strong airglow lines: Ly$\alpha$ and \ion{O}{1} 
$\lambda$1304.
\label{fig3}}
\end{figure}

\end{document}